\documentstyle[local]{article}
\newcommand{\beq}{\begin{equation}}
\newcommand{\eeq}{\end{equation}}

\begin{document}

\title {Astrometry with {\it Hubble Space Telescope}  (version 1.2)}

\section{Introduction}

\subsection{Astrometry with {\it Hubble Space Telescope}}
In 1990 NASA launched {\it Hubble Space Telescope}. In addition to cameras and spectrographs usable from the far ultraviolet to the near-infrared, the observatory contains three white-light interferometers. As part of engineering and science support their primary task was telescope guiding; to position and hold science targets within the science instrument apertures with tolerances approaching 0\farcs1, equivalent to 100 milliseconds of arc (100 mas). Pointing and tracking requires two such Fine Guidance Sensors (FGS), leaving the third free for astrometry. 

The design goal for astrometry with an FGS was 3 mas precision over the entire field of regard. When designed and built in the early 1980's, The FGS represented an order of magnitude improvement over existing ground-based techniques. {\it HST} launch delays provided sufficient time for ground-based techniques to equal and surpass this goal. Thus, our post-launch goal was redefined. This goal, 1 mas precision small-field astrometry, has been achieved, but not without significant challenges.

An {\it HST} FGS will remain a competitive astrometric tool for faint targets in crowded fields and for faint small-separation binaries, until the advent of large-aperture ground-based and longer-baseline space-based interferometers.

This article describes the science that can be done with the FGS and presents
some recent science results. We outline how  these data are modeled, acquired, and calibrated. We next show how the astrometer, FGS 3,  works as an interferometer. Finally, we present a guide to the literature that provides
additional detail for each item discussed in this article.

Astrometry with 1 mas precision is obtainable with the {\it HST} Wide Field Planetary Camera. The techniques are similar to those used in ground-based CCD
astrometry and should be  discussed separately.

\section{The choice of science targets}
The choice of {\it HST} astrometry science targets is rightly determined by what can be done from the ground. The unique capabilities of {\it HST} must remain reserved for projects demanding them. In this section we summarize the strengths of {\it HST} 
astrometry, pointing out
a few examples of the kinds of objects for which {\it HST} and the FGS are ideally 
suited. Since the FGS interferometer offers two observational modes, fringe scanning
and fringe tracking, we divide the remainder of this article by mode at each step.

\subsection{Fringe Scanning: Using {\it HST} for Binary Star Astrometry}
Four characteristics of a high-resolution ($<1"$) observing
technique must be considered when attempting to resolve a
target binary star: the source brightness limit, the resolving
capability, the astrometric accuracy, and the magnitude difference
between the components that can be observed.
Although all four parameters are interdependent, we
provide current limits for each characteristic
separately in order to illustrate the efficacy of
{\it HST} FGS3 as an astrometric instrument.

For binary stars  {\it HST} FGS3
observations provide results currently
unachievable with ground-based techniques. {\it HST} FGS3 is
accurate to 1-3 mas at a resolution limit of 15 mas for
magnitude differences less than 2 mag. It can effectively
observe targets to at least $V \sim15$. At component separations
greater than 200 mas, it can bridge brightness differences
of at least  $\Delta V  = 4$ mag, as demonstrated by the
detection of Gliese 623B.

Future developments in ground-based interferometers
should improve the accuracy of astrometric
measurements and provide better resolution, but will still be
limited to relatively bright targets and moderate magnitude
differences. {\it HST} FGS3 is the only high-resolution instrument
currently available that (1) can provide high-precision
astrometry for very close binaries, (2) allows relatively faint
targets to be observed, and (3) can bridge at least moderately
large magnitude differences between the components
in a binary system.
\subsection{Fringe Tracking: Using {\it HST} to Obtain Parallaxes}
{\it In the absence of distance, astrophysics lacks precision.}

Our demonstrated measurement precision for parallaxes is 0.5 mas or better, given six epochs of observation at maximum parallax factor. 
Often a parallax target is unmeasurable because reference stars are
not located within the field of view of the measuring device, especially true for small-format (typically several square arcmin) CCD cameras. An FGS 
provides a large field of regard (Fig. 1), shaped somewhat like a pickle. This shape comes from the the pick-off mirror (Fig. 5) in the {\it HST} focal plane.
Parallax observations are generally spaced by six months, during which time, due to {\it HST} solar array illumination constraints, the FGS field of view rotates by 180\arcdeg. The lens-shaped region (Fig. 6)
in common to the two extreme orientations provides a relatively large field  with
short axis = 3\farcm5 and long axis = 14'. 

FGS 3 has a large dynamic range, able to obtain fringe tracking position measures for stars in the magnitude range $4 \le V \le 17$. This large dynamic 
range is provided by a neutral density filter 
that reduces the magnitude of bright stars by 5 mag. The unfiltered range is $8.5 \le V \le 17$. 

{\it HST} can obtain precise parallaxes for binaries whose components are separated $0\farcs03 < \rho < 1\farcs0$. Nearly simultaneous fringe scanning and tracking  measurements provide the  component positions relative to reference stars, essential in determining the center of mass of a system, and thus individual component masses. 

Lastly there is timeliness of result. One no longer need wait 3-6 years for the parallax of astrophysically interesting objects. The distance to a sufficiently interesting and important object can be obtained on the same time scale as other astrophysical information. If an object or class of objects is interesting now, a theory can be tested now.

\section{Science Results}
Science targets have included nearby stars used as probes for an extrasolar planet search;
  stars whose astrophysics would be greatly aided by accurate distances;
 low-mass binary stars to define the mass-luminosity relation for the lower main sequence;
 members of the Hyades Cluster, key to the distance scale; extragalactic objects (QSO) to provide an inertial reference frame for {\it HIPPARCOS}; and the epitome of crowded fields, an extragalactic cluster, R136 in the Large Magellanic Cloud. 

Once again we divide by technique, discussing representative results from first fringe scanning, then fringe tracking.
\subsubsection{Fringe scanning and binary stars}
Fringe scanning science depends on deconvolution and the assumption that all objects in the field are point sources. 
For a recent science result we turn to the low-mass binary Wolf 1062 = Gl 748,
observed in support of a lower main sequence mass-luminosity project.
Fig. 2 presents fringe scans of this binary along the two orthogonal axes of FGS 3. Since each observed fringe is a linear superposition of two fringes (one for each component star in the binary), it is modeled with two identical single star fringes. Their placement along the axis and relative amplitudes provide separation and brightness differences. Relative separation along the two axes provides position angle information. Fig. 3 shows 
all the measured separations and position angles on the best-fit orbit. The mean absolute difference between the observed and computed separations is only 1.1 mas, and
only 0\fdg77 in the position angle. Residual 
vectors for all data points are drawn, but are smaller than
the points. Perhaps it is even more impressive to realize that
the box illustrated is only 0\farcs6 in size, so that good seeing 
from the ground would result in a stellar image the size of the
entire figure.

A combination of fringe scanning and fringe tracking observations for the low-mass binary, L722-22, has yielded  a relative parallax  $\pi = 0\farcs1656 \pm 0\farcs0008$ and component masses of 0.179 and 0.112 $M_{\sun} $ with formal random errors for the mass as low as 1.5\%.

\subsubsection{Fringe tracking: positions in a reference frame}
Fringe tracking science is primarily that of relating the position of a target to positions
of stars defining a reference frame. Here we discuss a Hyades parallax program and
a fundamental astrometry project involving {\it HIPPARCOS}.

Trigonometric parallax observations were obtained for seven Hyades members in six fields of view. These have been analyzed along with their proper motions to
determine the distance to the cluster. Formal uncertainties on individual parallaxes average 1 mas. This relatively large error is due both to poor spatial distribution and small number of reference stars in each field. Knowledge of the convergent point and mean proper motion of the Hyades
is critical to the derivation of the distance to the center of the cluster. Depending on the choice of the
proper-motion system, the derived cluster center distance varies by 9\%.  Therefore, a full utilization of the {\it HST} FGS parallaxes awaits the establishment
of an accurate and consistent proper-motion system.

Observations of separations of {\it HIPPARCOS} stars from extragalactic objects have been made to determine the rotation of the {\it HIPPARCOS} instrumental system with respect to the ICRS-VLBI reference frame. A determination from 78 observations yields accuracies on the order of 2 mas rms in the coordinate rotational offsets near the mean epoch of the {\it HST} observations and 2 mas yr$^{-1}$ in the coordinate rotation rates. The main contributing sources of error are {\it HST} measurement errors and proper motion errors introduced by the three year time difference between the mean {\it HIPPARCOS} and {\it HST} observational epochs.

\section{Data modeling}

Our two examples are a binary star for fringe scanning and a parallax field for fringe tracking.
\subsection{Fringe scanning}
We consider a binary star as the simplest object requiring fringe scanning.

We assume that the fringe produced by a binary star is a linear superposition of the fringes produced by two single stars. The interferometer response to an actual binary star is shown in Fig. 2. If $F(x)$ is the fringe produced by a single star on the x-axis, a binary star should be described by
\beq
D(x) = A \times F(x+z_{x}) + B\times F(x+z_{x}+S_x)
\eeq
where A and B are the relative intensities of the two components (constraining $A + B =1$). The zero point offset, $z_x$, and a component separation, $S_x$, complete the model, there being a similar expression for the fringe produced along the y-axis. $S_x$ and $S_y$ yield the binary separation, and, once transformed to equatorial coordinates, the position angle. We obtain the magnitude difference from
\beq
\Delta m = -2.5\times log(A/B)
\eeq

If a binary is observed in fringe tracking mode, the FGS will lock on an erroneous zero-crossing position which is generated by two
closely overlapping s-curves.  A fringe scanning mode observation
gives relative positions of the two components. One can determine the
relative positions
with an accuracy and precision of 1 mas, once the fringes from the two stars are deconvolved.

\subsection{Fringe tracking}
We consider a typical parallax target and associated reference stars
as an example of fringe tracking. The primary science target is ideally surrounded by 5 - 10 other stars used as a reference frame relative to which we determine position and motion. 

Fringe tracking astrometry is generally a two-step process. We first determine the characteristics of the reference frame. One of the
epochs of observation is chosen as the constraint plate.
From these data we determine the scale and rotation relative to the constraint plate  for each observation set within a single orbit.  Since for many of our targets the observation sets span over two years, we also include the effects of reference star parallax ($\pi$) and proper motion ($\mu$). 
\beq
\xi = ax + by + c -(P_x*\pi + \mu_x*t) 
\eeq 
\beq 
\eta = dx +ey +f- (P_y*\pi + \mu_y*t)
\eeq

The orientation relative to the celestial sphere is obtained from ground-based astrometry. Uncertainties in the field orientations are generally $0\fdg03 < \epsilon_\theta < 0\fdg09$. We obtain the parallax factors, $P_x$ and $P_y$ from a JPL Earth orbit predictor. 
Finally for a rich-enough reference field ($n>4$ stars) we constrain $\Sigma \mu = 0$ and   
$\Sigma \pi = 0$ for the entire reference frame. 

The second step consists of applying the plate constants to the measurements
of the science target. Plate coefficients $a, b, ..., f$ are applied
as constants, while we solve for science target $\pi$ and $\mu$ in the above equations.

\section{The astrometer}
\subsection{The design}
FGS 3 is an interferometer. Interference takes place in a prism that has been sliced in half,  had a quarter-wave retarding coating applied, and then reassembled. Fig. 4 shows one of these Koester's Prisms. Most of the FGS consists of supporting optics used to feed the Koester's Prisms (Fig 5). In particular the star selectors walk the 5" instantaneous field of view throughout the interferometer field of regard shown in Fig 1. The output of each face (face A and face B) is measured by a PMT. These signals are combined
\beq
S = {A-B \over A+B}
\eeq
to form a signal, S, that is zero for waves exactly vertically incident on the Koester's Prism front face. Tilting the wavefront back and forth (equivalent to pointing the telescope slightly off, then on target, then slightly off to the other side)
generates the fringe pattern seen in Fig. 2. A perfect instrument would generate a
perfectly symmetric fringe pattern. The significant spherical aberration of the as-built
{\it HST} primary mirror, in the presence of internal FGS misalignments, produces
a signature in the fringe which mimics coma. Coma causes decreased modulation and multiple
peaks and valleys in a fringe. FGS 3 produces the least complicated fringes over the
widest area within the field of regard. 

A replacement FGS installed in 1997 contains an articulated fold flat that removes most of the internal misalignments. This FGS (FGS 1r) produces nearly perfect
fringes and should at a minimum yield superior binary star results. FGS 3 will remain in servie until FGS 1r is fully calibrated in 2000.
\subsection{FGS operating modes}
We discuss strategies for obtaining the highest quality data possible. Our goal, 1 mas precision small-field astrometry, has been achieved, but not without significant challenges. These included a mechanically noisy on-orbit environment, the self-calibration of FGS 3, and significant temporal changes in our instrument. Solutions include a denser set of drift check stars for each science observation, fine-tuning exposure times, overlapping field observations and analyses for calibration, and a continuing series of trend-monitoring observations.

The single greatest contributor to data quality is to treat all data aquired in the
same orbit as a unit observation, e.g., a 'plate'.

\subsection{Fringe scanning}
The target is placed in the center of the pickle and the star selectors are commanded to move the instantaneous field of view across the target star image. This action produces a fringe. In practice 10 - 30 scans are obtained in 
a reciprocating pattern forwards and back across the star. The final fringe results from a reversal, shift and add process. We have evidence that over the span of an orbit the positions reported by FGS 1 and FGS 2 for the guide stars change. This results in a drift-like motion of FGS 3, the astrometer. Drift can exceed 30 mas over the span of 36 minutes. The x and y drift rates are generally dissimilar and the drift is not of constant rate. However, since a single scan across a science target requires about a minute, the drift per scan is reduced to less than 1 mas. The reciprocating data acquisition strategy is very nearly self-compensating for drift.

\subsection{Fringe tracking}

For fringe tracking, onboard electronics locate the zero crossing between the highest positive and lowest negative fringe peak (see Fig. 2). The position of this zero crossing is determined at a 40 Hz rate during an observation time ranging $10 < t < 300 s$. The median of $>2400$ zero crossings provides a robust position estimate. The star selectors are used to move the instantaneous field of view from one star to the next in the
FGS field of regard shown in Fig. 1. 

Drift is correctable, but imposes additional overhead, reducing the time available within an orbit to observe the science target. An astrometric observation set must contain visits to one or more reference stars, multiple times during each observation sequence. Presuming no motion intrinsic to all reference and science target stars over a span of 40 minutes, one determines drift and corrects the reference frame and target star for this drift. As a result we reduce the error budget contribution from drift to less than 1 mas. In Fig. 1, the science target might be observed three times during an observing session (single orbit) and each reference star twice.

\section{Astrometric Calibration}
The calibration of the fringe scanning mode has as its goal the determination of separation and magnitude differences between components of binary stars. As discussed earlier all fringe scanning targets are 
observed near the pickle center. In contrast fringe tracking takes place over the entire field of regard. Calibration requirements for these two modes of operation are quite different.
\subsection{Fringe scanning calibration}
The most fundamental calibration for any astrometric instrument is to determine the scale in units of the celestial sphere. The FGS scale comes from measurements of several  calibration  binary stars  frequently  observed  by  speckle interferometry with 4-m class telescopes.  On the basis of these speckle measures, extremely well-determined orbital elements were obtained. These binary star orbits yield, for any date of {\it HST} FGS observation, accurate angular separations of the binary, and hence, the scale.

Secondary calibration consists of the development of a library of suitable single star fringe templates. This library is required since fringe morphology weakly depends on star color. From a library of templates
the required number (one for each component of a binary or multiple star system) are chosen  and fit to the binary star fringe using the least-squares algorithm, GaussFit, that allows for errors in both independent and dependent variables. 

\subsection{Fringe tracking calibration}
The Optical Telescope Assembly (OTA) of the {\it HST} is an aplanatic Cassegrain telescope of Ritchey-Chr\`etien
  design.
Our initial problem included mapping and removing optical distortions whose effect on measured positions exceeded 0\farcs5. There was no existing star field with catalogued 1 mas precision astrometry, our desired performance goal, to use as a fiducial grid. Our solution was to use FGS3 to calibrate itself. As a result of this activity  distortions are reduced to better than 2 mas over much of the FGS3 field of regard (Fig. 1). This model is called the Optical Field Angle Distortion (OFAD) calibration.

To describe these distortions we have adopted a pre-launch functional form originally developed by the instrument builder, the Perkin-Elmer Corporation.
These distortions can be
 described (and modeled to the level of one millisecond of arc) by the two
dimensional fifth order polynomials: 
\begin{eqnarray}
x' = a_{00} + a_{10}x +a_{01}y + a_{20}x^{2} + a_{02}y^{2} + a_{11}xy 
+a_{30}x(x^{2}+y^{2}) + a_{21}x(x^{2}-y^{2}) \nonumber \\
+  a_{12}y(y^{2}-x^{2}) + a_{03}y(y^{2}+x^{2}) 
+a_{50}x(x^{2}+y^{2})^{2} + a_{41}y(y^{2}+x^{2})^{2}  \nonumber \\
+  a_{32}x(x^{4}-y^{4}) + a_{23}y(y^{4}-x^{4}) 
+a_{14}x(x^{2}-y^{2})^{2} + a_{05}y(y^{2}-x^{2})^{2}  \nonumber \\
\nonumber \\
y' = b_{00} + b_{10}x +b_{01}y + b_{20}x^{2} + b_{02}y^{2} + b_{11}xy 
+b_{30}x(x^{2}+y^{2}) + b_{21}x(x^{2}-y^{2}) \nonumber \\
+ b_{12}y((y^{2}-x^{2}) + b_{03}y(y^{2}+x^{2}) 
+b_{50}x(x^{2}+y^{2})^{2} + b_{41}y(y^{2}+x^{2})^{2}  \nonumber \\
+ b_{32}x((x^{4}-y^{4}) + b_{23}y(y^{4}-x^{4}) 
+b_{14}x(x^{2}-y^{2})^{2} + b_{05}y(y^{2}-x^{2})^{2} \nonumber\\
\end{eqnarray}
where x, y are the observed position 
within the FGS field of view, x', y' 
are the corrected position, and the numerical values
of the coefficients $a_{ij}$ and $b_{ij}$ are determined by calibration. 
Ray-traces were used for the initial estimation of the
OFAD.
Gravity release, outgassing of graphite-epoxy structures within the FGS, and
post-launch adjustment of the {\it HST} secondary mirror required that 
the final determination of the OFAD coefficients $a_{ij}$ and $b_{ij}$ be
made by an on-orbit calibration. 

M35 was chosen  as the calibration field.  Since the ground-based
positions of our target calibration stars were 
known only to 23 milliseconds of arc, the positions of the stars 
were estimated 
simultaneously with the distortion parameters. This was
accomplished during a marathon calibration, executed 
on 10 January 1993 in  FGS 3.   The entire 19 orbit sequence is shown in Fig. 6.
GaussFit 
was used  to simultaneously estimate the relative star positions, the 
pointing and roll of the telescope during each orbit (by quaternions), 
the magnification of
the telescope, the OFAD polynomial coefficients, and four 
parameters that describe the star selector optics inside the
FGS.

\subsection{Other fringe tracking calibrations}

Because each FGS contains refractive elements (Bradley et al. 
1991), it is possible that the position measured for a star could
depend on its intrinsic color. Changes in position would depend on star color, 
but the direction of shift is expected to be
constant, relative to the FGS axes. This lateral color shift would be 
unimportant, as long as target and reference stars have
similar color. However, this is certainly not the case for many of the science targets
(e.g., very red stars such as  Proxima Centauri and Barnard's Star). 

Finally, to provide the large dynamic range, a neutral-density filter can be placed in front of the Koester's Prism. As a consequence there will be a small but calibratable shift in position (due to filter wedge) when comparing 
the positions of the bright star to the faint reference frame.  The shift is constant in 
direction (relative to FGS 3) and size, since the filter does not rotate within 
its holder.

\subsection{Maintaining the fringe tracking calibration}
The FGS3 graphite-epoxy optical bench was predicted to outgas for a period of time after the launch of {\it HST}. The outgassing was predicted to change the relative positions of optical components on the optical bench.  The result of whatever changes were taking place was a change in scale.  The amount of scale change was far too large to be due to true magnification changes in the {\it HST} optical assembly. Two of the parameters
that descibe the star selector optics
 cause a scale-like change, if they are allowed to vary with time. 

The solution was to revisit the M35 calibration field periodically to monitor these scale-like changes and other slowly varying non-linearities. Revisits will be required as long as it is desirable to do 1 mas precision astrometry with any FGS. The result of this activity judges the validity of the
current OFAD coefficients and provides warning for the need for recalibration.
With these data we remove the slowly varying component of the OFAD, so that uncorrected distortions remain below 2 mas for center of FGS 3. The character of these changes is 
 generally monotonic with abrupt jumps in conjunction with {\it HST} servicing missions.

\section{ Bibliography}

  Here are pointers to a number of
key papers in the development of {\it HST} interferometric astrometry, and to some recent results.

\bigskip

A detailed description of the FGS interferometer can be found in

{\small The flight hardware and ground system for Hubble Space Telescope astrometry. Bradley, A., Abramowicz-Reed, L., Story, D., Benedict, G. \& Jefferys, W. 1991, \pasp, 103, 317 }

\normalsize and in
 
{\small Lupie, O., Nelan, E. P., 1998, FGS Instrument Handbook, version 7, (STScI)
\bigskip

\normalsize Our choice of astrometer (there being three from which to choose) is documented in

{\small Astrometric performance characteristics of the Hubble Space Telescope fine guidance sensors. Benedict, G. F., et al. 1992, \pasp, 104, 958}
 
\bigskip

\normalsize Our primary calibration tool is GaussFit, a least-squares and robust estimation
algorithm described in 

{\small Jefferys, W., Fitzpatrick, J., and McArthur, B.  1988 Celest. 
Mech. 41, 39.}

\bigskip

\normalsize Calibration and observing strategies for fringe scanning are discussed in

{\small Binary star observations with the Hubble Space Telescope Fine Guidance Sensors. I - ADS 11300. Franz, O. G., et al. 1991, \apjl, 377, L17} 

\normalsize and

 {\small A deconvolution technique for Hubble Space Telescope FGS fringe analysis. Hershey, J. L. 1992, \pasp, 104, 592}
 
\bigskip

\normalsize Introductions to  deriving and maintaining the fringe tracking calibration and observing strategies include

{ \small Astrometry with Hubble Space Telescope Fine Guidance Sensor number 3: Position-mode stability and precision. Benedict, G. F., et al. 1994, \pasp, 106, 327},

{\small Maintaining the FGS3 Optical Field Angle Distortion Calibration. McArthur, B., Benedict, G. F., Jefferys, W. H. \& Nelan, E. 1997, The 1997 HST Calibration Workshop with a new generation of instruments. Editors Stefano Casertano, Robert Jedrzejewski, Charles D. Keyes, and Mark Stevens. Baltimore, MD : Space Telescope Science Institute 1997, p. 472. }

 \normalsize and

{\small Working with a space-based optical interferometer: HST Fine Guidance Sensor 
3 small-field astrometry.  Benedict, G. F., et al. 
1998, \procspie, 3350, 229} 

\bigskip

\normalsize Recent astrometric results naturally divide into fringe scanning and fringe tracking. For fringe scanning, 

{ \small Astrometric Companions Detected at Visible 
Wavelengths with the Hubble Space Telescope Fine Guidance Sensors. 
Franz, O. 
et al. 1994, 
American Astronomical Society Meeting, 185, 85.24 }

\normalsize discusses the first results for large $\Delta m$.

 The first definitive binary orbit determined with the Hubble Space 
Telescope Fine Guidance Sensors was for Wolf 1062 = Gliese 748 

{\small Franz, O. G., et al. 1998, \aj, 116, 1432}

\normalsize From the beginning, stellar interferometers have explored stellar diameters.
The FGS has determined interferometric angular diameters of Mira variables 

{\small Lattanzi, M. G., Munari, U., Whitelock, P. 
A. \& Feast, M. W. 1997, \apj, 485, 328 }.

\normalsize FGS fringe scanning has contributed to the optical Mass-Luminosity Relation at the lower Main Sequence (.08 to .20 $M_{\sun} $), discussed in

{\small Henry Henry, T.\ J. , Franz, O. G.,
Wasserman, L. H., Benedict, G. F., Shelus, P.J., Ianna, P.A.,
Kirkpatrick, J. D., \& McCarthy, D. W. 1999, ApJ, 512, 864}
\bigskip

\normalsize Recent fringe tracking results include
a report on the search for planets around Proxima Centauri

{\small Benedict, G. F., McArthur, B.,
Chappell, D. W., Nelan, E., Jefferys, W. H., Van Altena, W., Lee, J.,
Cornell, D., Shelus, P. J., Hemenway, P. D., Franz, O. G., Wasserman,
L. H., Duncombe, R. L., Story, D., Whipple, A., \& Fredrick,
L. W. 1999, AJ, 118, 1086},

\normalsize parallaxes of two high-velocity stars 

{\small Macconnell, D. J., Osborn, W. H. \& Miller, R. J. 1997, \aj, 114, 1268}, 

\normalsize and the distance to the Hyades Cluster based on FGS parallaxes 

{\small Van Altena, W. F.
et al., 1997. \apjl, 486, L123}.

\normalsize An FGS program to link the {\it HIPPARCOS} reference frame to an extragalactic 
reference system  is detailed in 

{\small Hemenway, et al 1997, \aj, 114, 2796}

\bigskip

\normalsize Finally a combination of fringe tracking and scanning yielded a parallax and component masses for the low-mass 
binary L722-22 

{\small Hershey, J. L. \& Taff, L. G. 1998, \aj, 116, 1440}

\bigskip

\normalsize A snapshot, circa 1994, of the scientific capabilities of the FGS can be found in

Astronomical and astrophysical objectives of sub-milliarcsecond optical astrometry:
proceedings of the 166th Symposium of the International Astronomical Union held in 
the Hague, Netherlands, August 15-19, 1994 Edited by E. Hog, P. Kenneth Seidelmann,
International Astronomical Union. Symposium no. 166, Kluwer Academic, Publishers,
Dordrecht, 1995

Papers include

{\small 
Fringe Interferometry in Space: The Fine Structure of R136A with the 
Astrometer  Guidance Sensor Aboard HST. Lattanzi M. G. et al., pg 95 

\normalsize and

{\small Hubble Space Telescope: A Generator of Submilliarcsecond Precision Parallaxes. Benedict, G. F., et al. , pg. 89 }}
\bigskip

\hspace{3in}George F. (Fritz) Benedict
\bigskip

This article relied heavily on inputs from and reviews by B. McArthur, E. Nelan, O. Franz, L. Wasserman, and W. Jefferys.

\clearpage

\begin{figure}
\epsscale{0.7}
\plotone{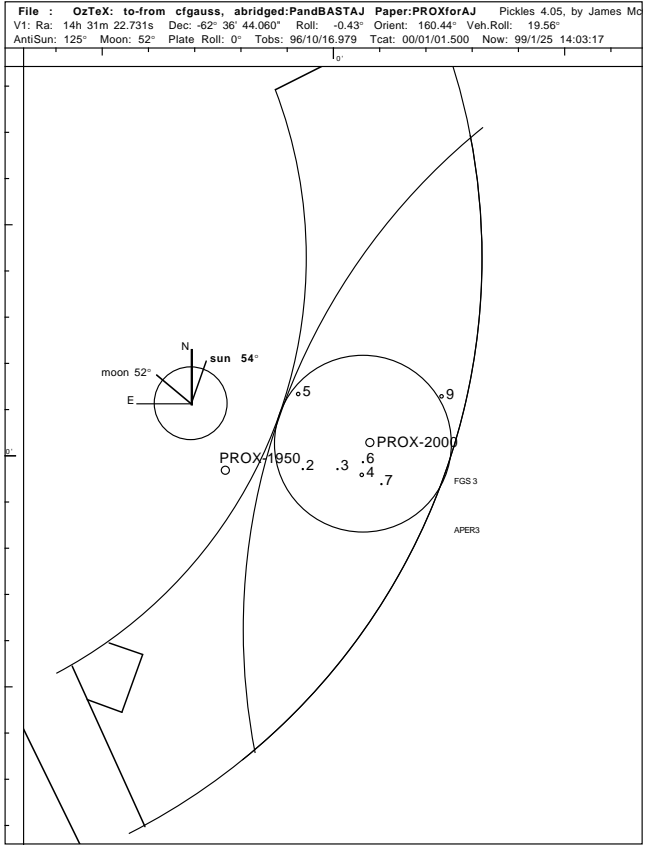}
\caption{ Field of Regard of FGS 3 on the sky at the location
of the planet search target, Proxima Centauri on 16 October 1996. The position of
Proxima Centauri at two epochs is plotted. Also shown are the reference stars,
relative to which parallax and proper motion are obtained. Tick marks are separated by 1 arc minute.} \label{fig-1}
\end{figure}
\clearpage

\begin{figure}
\epsscale{1.0}
\epsscale{0.8}
\plotone{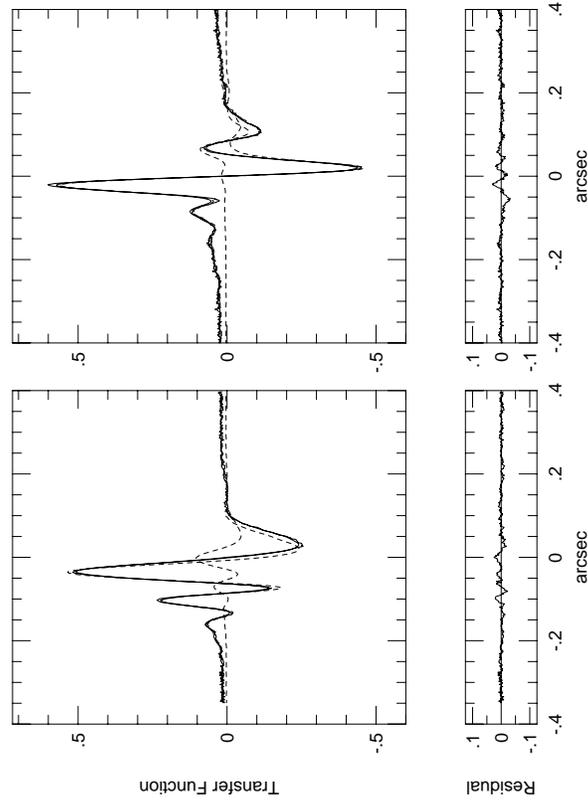}
\caption{ Analysis of the 1995.5712 observation of Wolf
1062.  The computed (smoothed solid) curves in X (bottom) and Y (top)
represent best-fitting linear superpositions of two single-star fringes
(dashed curves).  The relative displacements
of the single-star curves are 0\farcs0329 $\pm$ 0\farcs0003 and  
0\farcs0862 $\pm$ 0\farcs0002 along FGS X and Y respectively.
They yield separation and position angle, while the relative amplitudes
of the single-star curves yield a magnitude difference for the binary
components. The systematic trends in the curve-fitting residuals
(bottom panels) are due to imperfect single-star calibrations.  Their
effect upon the astrometry and photometry of the Wolf 1062 components
is negligible.} \label{fig-2}
\end{figure}
\clearpage

\begin{figure}
%\epsscale{1.0}
\epsscale{0.8}
\plotone{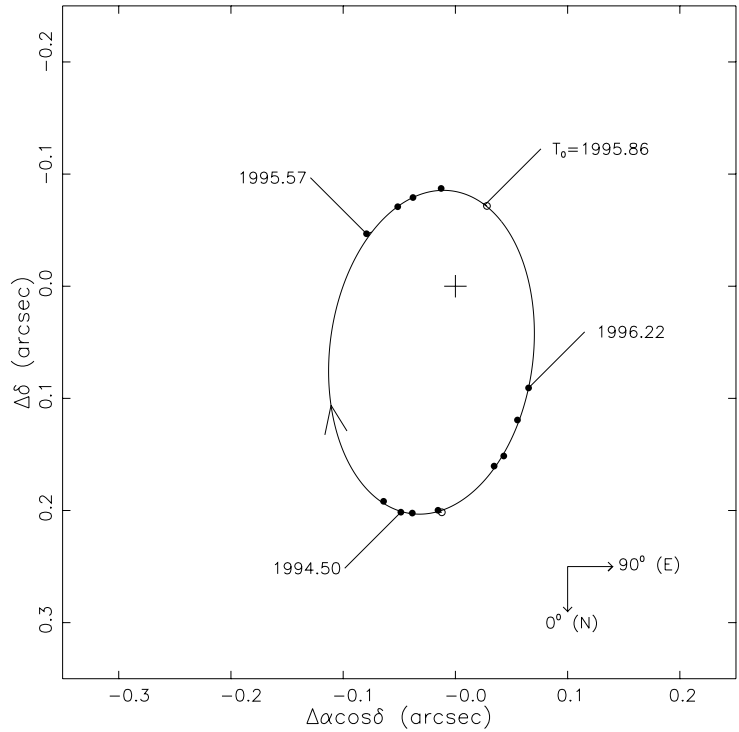}
\bigskip
\bigskip
\caption{ {\it HST} FGS3 measurements of Wolf 1062AB and the orbital path predicted by the derived elements  are shown. Three 
epochs of observation and the time of periastron are labelled. Residual vectors for all data points are plotted, but are smaller than the points themselves.  } \label{fig-3}
\end{figure}
\clearpage

\begin{figure}
\epsscale{1.0}
\epsscale{0.8}
\plotone{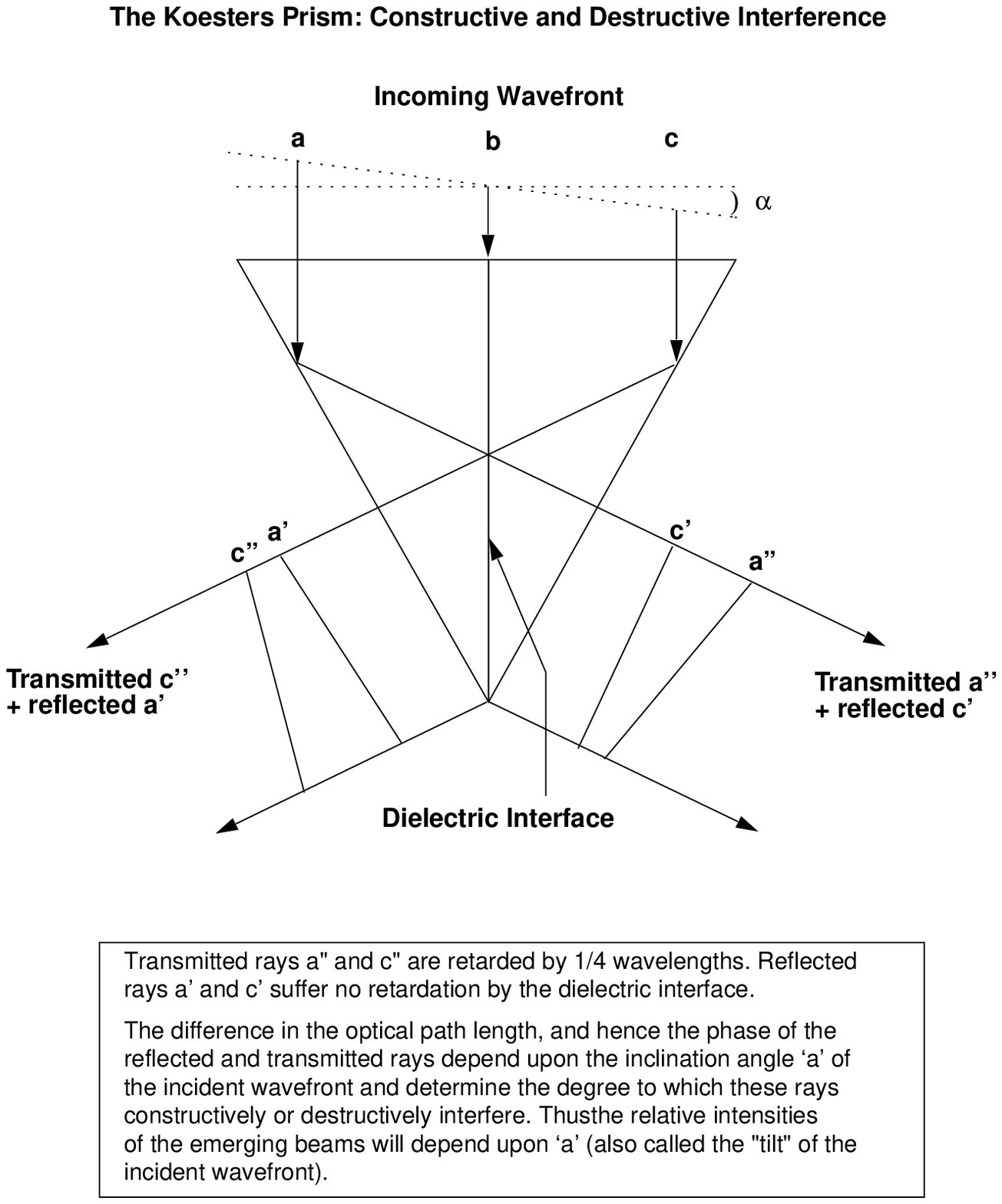}
\caption{ The heart of the FGS interferometer, the Koester's Prism.  } \label{fig-4}
\end{figure}
\clearpage
\begin{figure}
\epsscale{0.8}
\plotone{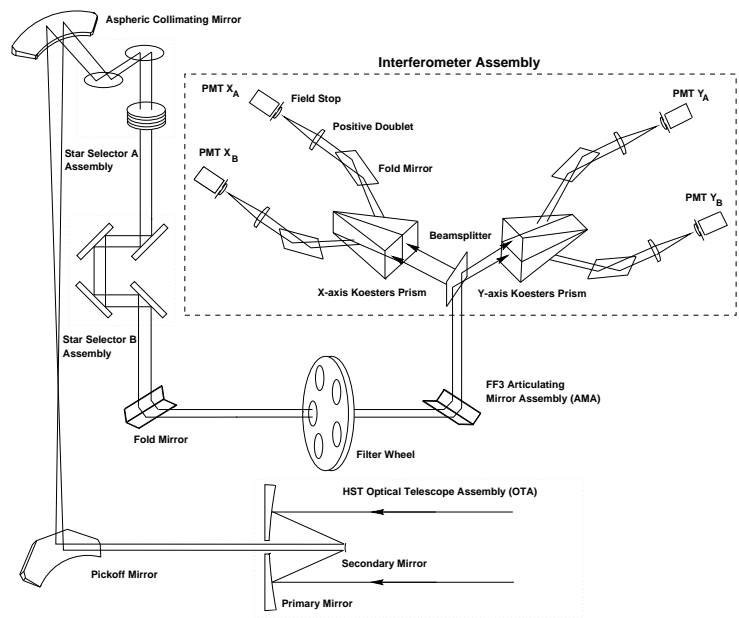}
\caption{ The optical layout of a Fine Guidance Sensor. } \label{fig-5}
\end{figure}
\clearpage
\begin{figure}
\epsscale{0.8}
\plotone{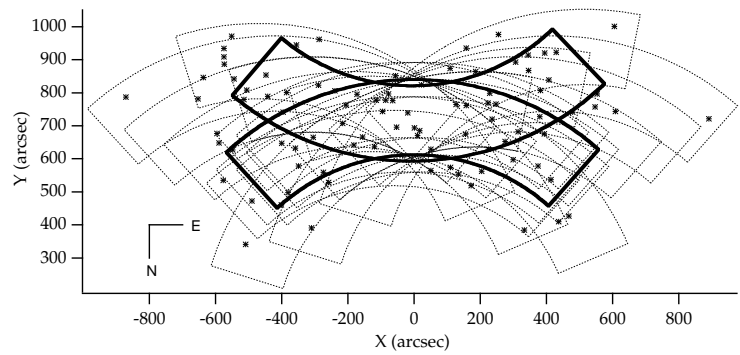}
\caption{ The fringe tracking self-calibration of FGS 3. Stars in the cluster M35 were observed at nineteen different pointings and rolls of {\it HST}. The pickle-shaped
regions denote the total FGS field of regard. The black dots are stars within the cluster M35. The lens-shaped area
in the middle shows the field of view for parallax studies. } \label{fig-6}
\end{figure}
\clearpage

\end{document}